\documentstyle[aps,preprint,eqsecnum]{revtex}
\begin{document}
\draft \tighten
%%
%\tighten
%@twocolumnfalse\endcsname
\title{Strengths of singularities in spherical symmetry.}
\author{Brien C. Nolan\footnote{e-mail: nolanb@ccmail.dcu.ie}}
\address{School of Mathematical Sciences,\\ Dublin City University,\\
Glasnevin, Dublin 9,\\ Ireland.}
%\date{\today}
\maketitle
%\newpage
\begin{abstract}
Covariant equations characterizing the strength of a singularity in
spherical symmetry are derived and several models are
investigated. The difference between central and non-central
singularities is emphasised. A slight modification to the
definition of singularity strength is suggested. The gravitational
weakness of shell crossing singularities in collapsing spherical
dust is proven for timelike geodesics, closing a gap in the proof.
\newline
\end{abstract}

\section{Introduction}

Over thirty years have passed since the appearance of the first of
a series of theorems establishing that under very general
circumstances, space-times must develop singularities. This first
result, due to Penrose \cite{pen65}, appeared in 1965 and the body
of work which grew up around the singularity theorems is contained
in the book of Hawking and Ellis \cite{HE} which was first
published in 1973. However, out understanding of the nature of
these singularities remains far from complete. This lack is best
exemplified by the absence of a proof, or definitive refutation,
of the cosmic censorship conjecture (CCC)
\cite{waldrev,clarkerev}.

An important aspect of a singularity is its gravitational strength
\cite{tipler}. A singularity is termed gravitationally strong, or
simply strong, if it destroys by crushing or stretching any object
which falls into it. The most familiar example is the singularity
at $r=0$ in the Schwarzschild solution. (Throughout this paper, we
will refer to a singularity at $r=0$ as a central singularity, and
to others as non-central.) A radially infalling object
is infinitely stretched in the radial direction and crushed in the
tangential directions, with the net result of crushing to zero
volume. A singularity is termed weak if {\em no} object which
falls into the singularity is destroyed in this way. The
mathematical description of these ideas runs as follows
\cite{tipler,clarkerev}.

Let $\gamma:[\tau_0,0)\rightarrow M$ be a causal geodesic which
approaches a singularity as $\tau\rightarrow 0^-$. Define
$J_\nu(\gamma)$ for $\nu\in[\tau_0,0)$ to be the set of maps
$Z:[\nu,0)\rightarrow TM$ satisfying the geodesic deviation
equation along $\gamma$ such that $Z(\tau)\in T_{\gamma(\tau)}M$,
$g_{ab}k^a(\tau)Z^b(\tau)=0$ where $k^a$ is the tangent to
$\gamma$ and
\[ Z(\nu)=0,\]
i.e. $J_\nu(\gamma)$ is the set of Jacobi fields along $\gamma$
which vanish at $\gamma(\nu)$. Along a timelike geodesic, three
independent Jacobi fields define, via the exterior product, a
volume element $V(\tau)$ along $\gamma$. Along a null geodesic,
two such fields define an area element which we also denote
$V(\tau)$. A timelike (null) geodesic terminates in a {\em strong
curvature singularity} if for all $\nu\in[\tau_0,0)$ and all
independent triads (dyads) in $J_\nu(\gamma)$ we have
\[ \lim_{\tau\rightarrow 0^-} V(\tau)=0.\]
Then the singularity itself is said to be {\em strong} if every
causal geodesic which approaches it terminates in a strong
singularity. The geodesic terminates in a weak singularity if the
limit above is finite and non-zero, and the singularity is weak if
every causal geodesic approaching it terminates weakly. We will argue below
for a slight modification of this definition whereby the term
strong is also attached to a singularity if the norms of the
Jacobi fields themselves have zero or infinite limit.

The importance of the notion of the gravitational strength of a
singularity for the CCC is that a statement of such possibly need
not rule out the occurrence of naked weak singularities. This is
based on the belief that one may extend the geometry of space-time
through a weak singularity without traumatic effects
\cite{joshi,ori-burko}. A general description of this extension
does not exist - indeed as far as the author can determine, only
two examples of this procedure exist in the literature, one due to
Papapetrou and Hamoui \cite{papa} and the other due to Clarke and
O'Donnell \cite{cod}. Both deal with extending through a shell
crossing singularity in collapsing spherical dust. However the
fact that at a weak singularity one has, along any timelike
geodesic, a finite non-degenerate triad of Jacobi fields, from
which it may be possible to construct a metric in a canonical way,
lends support to the idea.

Our aim here is to give a comprehensive analysis of the strengths
of singularities in spherical symmetry. This has been the arena of
some of the most interesting developments in general relativity in
recent years, and an understanding of what can and cannot occur in
spherical symmetry may be a valuable guide for more general
situations. Specifically, we study the Jacobi equations for
arbitrary radial causal geodesics. This allows us to give
covariant equations identifying the geometrical terms which
control the strength of the singularity. As one would expect, the
results are simpler than those obtained by Clarke and Krolak
\cite{CK} which apply to general space-times. We study three
different models which help illustrate the different situations
which occur, and by way of application, demonstrate that (i) a
non-central singularity is always weak along null directions and
(ii) the shell-crossing singularities in collapsing spherical dust
are weak. (This has only been demonstrated previously for radial
null directions \cite{newman}; we complete the proof by showing
that it is also true for radial timelike directions.)

\section{Radial causal geodesics, Jacobi fields and the volume element.}

The line element of a spherically symmetric space-time may be
written as
\begin{eqnarray}
ds^2&=&-2e^{-2f}du\,dv+r^2d\omega^2,\label{eq1}\end{eqnarray}
where $f=f(u,v)$, $r=r(u,v)$ and $d\omega^2$ is the line element
of the unit 2-sphere. The function $r$ is an invariant of the
space-time which we will, quite properly, refer to as the radius.
The coordinates $u$ and $v$ are both null, labelling the null
hypersurfaces generated by the two families of null geodesics
orthogonal to the orbits of the $SO(3)$ symmetry group of the
space-time. The form (\ref{eq1}) is invariant under the
transformations $u\rightarrow u^\prime(u)$, $v\rightarrow
v^\prime(v)$.

The non-vanishing Ricci tensor components are
\begin{mathletters}\label{eq2}
\begin{eqnarray}
R_{uu}&=&-2r^{-1}(r_{uu}+2r_uf_u)\label{eq2a}\\
R_{vv}&=&-2r^{-1}(r_{vv}+2r_vf_v)\label{eq2b}\\
R_{uv}&=&-2r^{-1}(r_{uv}-rf_{uv})\label{eq2c}\\
R_{\theta\theta}&=&\csc^2\theta R_{\phi\phi}=
1+2e^{2f}(r_ur_v+rr_{uv}).\label{eq2d}
\end{eqnarray}
\end{mathletters}

We use the convention that subscripts attached to lower case
letters refer to partial derivatives, but elsewhere refer to
tensor components in the associated coordinate basis. The
Misner-Sharp energy \cite{hayward} is
\begin{eqnarray}
E&=&\frac{r}{2}(1+2e^{2f}r_ur_v),\label{eq3}\end{eqnarray} and the
Weyl tensor is completely determined by the Newman-Penrose term
$\Psi_2$, calculated on a null tetrad based on the principal null
directions of the space time. Thus $\Psi_2$ is an invariant of the
space-time and is given by
\begin{eqnarray}
\Psi_2&=&\frac{e^{2f}}{3r}(r_{uv}+rf_{uv})-\frac{E}{3r^3}.
\label{eq4}\end{eqnarray} We note the further invariant, \begin{eqnarray}
e^{2f}f_{uv}&=&\frac{E}{r^3}+2\Psi_2-\frac{R}{12}, \label{eq5}\end{eqnarray}
where $R$ is the Ricci scalar.

The radial geodesic equations are
\begin{mathletters}\label{eq6}
\begin{eqnarray} -2e^{-2f}{\dot u}{\dot v}&=&\epsilon \label{eq6a}\\ {\ddot
u}-2f_u{\dot u}^2&=&0 \label{eq6b}\\ {\ddot v}-2f_v{\dot v}^2&=&0
\label{eq6c} \end{eqnarray}
\end{mathletters}
where $\epsilon=0$ for null geodesics and $-1$ for timelike; the
overdot is respectively, differentiation with respect to affine
parameter and proper time. Space-like geodesics will not concern
us here.

We now look at the Jacobi fields along arbitrary radial causal
geodesics, beginning with time-like geodesics.

The unit tangent to an arbitrary time-like curve $\gamma(\tau)$ in
the radial 2-space can be written in the form
\[ {\vec k}=h\frac{\partial}{\partial
u}+\frac{1}{2}e^{2f}h^{-1}\frac{\partial}{\partial v},\] where
$h=h(u,v)$. The condition that $\gamma$ be geodesic is then \begin{eqnarray}
h_v+2h^2e^{-2f}(h_u-2f_uh)&=0.\label{eq7}\end{eqnarray}This follows from the
geodesic equations (\ref{eq6}). The variation of any scalar
quantity $s$ along this geodesic is given by
\begin{eqnarray}
{\dot
s}&=&k^a\nabla_as=hs_u+\frac{1}{2}e^{2f}h^{-1}s_v,\label{eq8}\end{eqnarray}
and using (\ref{eq7}), \begin{eqnarray} {\ddot
s}&=&h^2(s_{uu}+2f_us_u)+\frac{1}{4}e^{4f}h^{-2}(s_{vv}+2f_vs_v)
+e^{2f}s_{uv}.\label{eq9}\end{eqnarray}

A Jacobi field $Z^a$ along $\gamma$ satisfies the geodesic
deviation equation \begin{eqnarray} {\ddot
Z^a}+{R_{cbd}}^aZ^bk^ck^d&=&0,\label{eq10}\end{eqnarray}which is a linear
equation for $Z^a$ and so a basis for the Jacobi fields may be
found by obtaining all independent Jacobi fields in the radial
2-space and in the tangential 2-space.

We take
\[ {\vec \xi}_{(1)}=x(u,v)\frac{\partial}{\partial \theta}\qquad
{\vec \xi}_{(2)}=y(u,v)\csc\theta\frac{\partial}{\partial \phi}\]
as candidates for the Jacobi fields in the tangential 2-space.
Note that the norms of $\xi^a_{(1)}$ and $\xi^a_{(2)}$ are $|x|$
and $|y|$ respectively. The geodesic deviation equation
(\ref{eq10}) applied to $\xi^a_{(1)}$ yields the following
equation for $x$ (the same result applies to $y$):
\begin{eqnarray*}
r(4h^4(x_{uu}+2f_ux_u)+e^{4f}(x_{vv}+2f_vx_v)+4h^2e^{2f}x_{uv})&&\\
+2(2h^2x_u+e^{2f}x_v)(2h^2r_u+e^{2f}r_v)&=&0. \end{eqnarray*}
Using (\ref{eq8}) and (\ref{eq9}), this assumes the remarkably
simple form \begin{eqnarray} r{\ddot x}+2{\dot r}{\dot x}&=&0, \label{eq11}\end{eqnarray}
which can be integrated to obtain \begin{mathletters} \label{eq12}
\begin{eqnarray}
x(\tau)&=&x_0\int_{\tau_1}^\tau\frac{d\tau^\prime}{r^2(\tau^\prime)}
, \label{eq12a}\end{eqnarray}where $x_0$ is constant and we have included
the initial condition $x(\tau_1)=0$, so that ${\vec
\xi}_{(1)}(\tau_1)=0$. The second linearly independent solution is
$x(\tau)\equiv 0$. We obtain the same result for ${\vec
\xi}_{(2)}$:
\begin{eqnarray}
y(\tau)&=&y_0\int_{\tau_1}^\tau\frac{d\tau^\prime}{r^2(\tau^\prime)}
. \label{eq12b}\end{eqnarray}
\end{mathletters}

Before dealing with the radial Jacobi fields along the time-like
geodesics, we describe the situation for radial null geodesics. It
turns out to be remarkably simple. From (\ref{eq6a}), either
${\dot u}$ or ${\dot v}$ vanishes along a radial null geodesic.
Take it to be the latter. We can then integrate (\ref{eq6b}) to
obtain ${\dot u}=ce^{2f}$ and so the tangent is
\[ {\vec k}=ce^{2f}\frac{\partial}{\partial u}.\]
The variation of a scalar $s$ along the geodesic is ${\dot
s}=ce^{2f}s_u$ and
\[ {\ddot s}=c^2e^{4f}(s_{uu}+2f_us_u).\]
In the null case, there are only tangential Jacobi fields.
Candidates for such are given by ${\vec \xi}_{(1,2)}$ as above,
and it turns out that the norms $x,y$ obey the same equation as in
the time-like case, and thus the solutions are given by
(\ref{eq12a}).

We now treat the radial Jacobi fields along a time-like geodesic.
A space-like vector in the radial 2-space orthogonal to $k^a$ has
the form
\[ {\vec \xi}=ah\frac{\partial}{\partial
u}-\frac{1}{2}e^{2f}ah^{-1}\frac{\partial}{\partial v},\] where
$a=a(u,v)$. This has norm $|a|$. Using (\ref{eq9}), we find that
the condition for ${\vec \xi}$ to satisfy the geodesic deviation
equation along $k^a$ is \begin{eqnarray} {\ddot
a}+2e^{2f}f_{uv}a&=&0.\label{eq12c}\end{eqnarray}
Recall that according to
(\ref{eq5}), $e^{2f}f_{uv}$ is an invariant of the space-time.
Thus equations (\ref{eq12}) and (\ref{eq12c}) provide a {\em covariant}
set of equations which will determine the strength of the
singularity. To see how this comes about, we obtain the
relationship between $V(\tau)$ and the quantities $a,x,y$.

Since each of $a,x$ and $y$ satisfy second order linear ordinary
differential equations, there are six independent Jacobi fields
along a time-like geodesic $\gamma$. An arbitrary triad of
corresponding 1-forms is given by
\[ {\bf z}_\alpha=a_\alpha{\bf
e}+x_\alpha r^2d\theta+y_\alpha\sin\theta d\phi,\] where ${\bf
e}=(2h)^{-1}du-e^{-2f}hdv$ and $\alpha=1,2,3$. Then the general
volume element along $\gamma$ has the form \begin{eqnarray*}
V(\tau)&=&{\bf z}_1\wedge{\bf z}_2\wedge{\bf z}_3\\
&=&6a_{[1}x_{2}y_{3]}r^4\sin\theta{\bf e}\wedge d\theta\wedge
d\phi.
\end{eqnarray*}
The norm $\|{\bf W}\|$ of a $p-$form ${\bf
W}=W_{[{i_1}...{i_p}]}dx^{i_1}\wedge...\wedge dx^{i_p}$ is given
by
\[\|{\bf W}\|^2=W_{|{i_1}...{i_p}|}W^{{i_1}...{i_p}},\]
where the vertical bars indicate summation only over
$i_1<i_2<...<i_p$. This gives \begin{eqnarray}
\|V(\tau)\|&=&6|a_{[1}x_2y_{3]}|r^2. \label{eq13}\end{eqnarray}

The existence of six independent solutions of the geodesic
deviation equation indicates that in a general space-time, the
different $V(\tau)$ form a six parameter family. This surfeit of
possibilities would produce problems if one wanted to make
statements about singularity strengths based on the behaviour of
all such $V(\tau)$. However the definition of $J_{\tau_1}$ reduces
the number significantly. Note that ${\bf z}_\alpha$ vanishes if
and only if each of $a_\alpha, x_\alpha$ and $y_\alpha$ vanish.
The general solution for $a$ has the form
\[ a=c_+a_+ + c_-a_- \]
where $c_\pm$ are arbitrary constants and $a_\pm$ are any two
independent solutions of (\ref{eq12c}). A similar result holds for
$x$ and $y$. The initial condition $a(\tau_1)=0$ fixes the ratio
$c_+/c_-$, so that for the problem in hand, there is only {\em
one} choice, up to a constant multiple, for each of $a,x$ and $y$.
Therefore the norm of every relevant volume element has the simple
form \begin{eqnarray} \|V(\tau)\|&=&|axy|r^2\label{eq14}\end{eqnarray}where $a,x$ and $y$
here represent the {\em general} solutions of the appropriate
equations (\ref{eq12}) and (\ref{eq12c}), into which constants may
be absorbed.

We thus have a simple and direct way of assessing the strength of
a singularity. We determine the limiting behaviour of solutions of
(\ref{eq12}) and (\ref{eq12c}) as the singularity is approached
and then use (\ref{eq14}) to calculate $\lim_{\tau\rightarrow
0^-}\|V(\tau)\|$.

\section{Possible solutions of the main equations}

We now consider the various possible limiting behaviours which can
occur for solutions of the main equations, i.e. (\ref{eq12}) and
(\ref{eq12c}).

From (\ref{eq14}), the forms $rx$ and $ry$ arise naturally and
occur in the following proposition which applies to both $x$ and
$y$.

{\bf Proposition One} {\em (i) For a non-central singularity,
$\lim_{\tau\rightarrow 0^-} rx$ is finite and non-zero.
\newline
(ii) For a central singularity, if
\[ \lim_{\tau\rightarrow 0^-} \int_{\tau_1}^\tau
r^{-2}(\tau^\prime)\,d\tau^\prime < \infty,\] then
$\lim_{\tau\rightarrow 0^-} rx =0$. Otherwise
\[ \lim_{\tau\rightarrow 0^-} rx = -\lim_{\tau\rightarrow
0^-}\frac{1}{{\dot r}(\tau)}. \]}

Here and below, the limit refers to the limit as $\tau\rightarrow
0^-$ along a geodesic which approaches the singularity at
$\tau=0$. The proofs of part (i) and the first part of part (ii)
follow immediately from (\ref{eq12a}), and that of the second part
from (\ref{eq12a}) and l'Hopital's rule. A consequence of this is
that the strength of a non-central singularity is completely
determined by the limiting behaviour of $a$ at the singularity.

Next we summarise the possible behaviour of $a$ in the appropriate
limit. Define $F(\tau)=2e^{2f}f_{uv}(\tau)$ and
$G(\tau)=\tau^2F(\tau)$. Then (\ref{eq12c}) is equivalent to \begin{eqnarray}
\tau^2{\ddot a}+G(\tau)a&=&0,\label{eq15}\end{eqnarray}and we wish to
determine the behaviour of $a$ at $\tau=0$ which is a singular
point of (\ref{eq15}). We quote the following results from Bender
and Orszag \cite{bo}, which should also be found in any text on
linear differential equations. All the asymptotic relations below
hold as $\tau\rightarrow0^-$ and $c,c_\pm$ are arbitrary
constants.

The equation is said to have a {\em regular singular point} at $\tau=0$ if
$G(\tau)$ is analytic in a neighbourhood of $\tau=0$. Otherwise, $\tau=0$ is called an {\em irregular singular point}.
For a regular singular point, we define $G_0=G(0)$. In fact the results below apply more generally, namely if $G(\tau)=O(1)$ as $\tau\rightarrow 0^-$ with a very low degree of differentiability; $G\in C^1(-\tau_0,0]$ is sufficient. The method of Frobenius
applies. The roots of the indicial equation are
\[ v_{1,2}=\frac{1}{2}\pm\frac{1}{2}(1-4G_0)^{1/2}.\]
The following possibilities arise.
\newline
{\bf (RSP1)} $v_1-v_2 \not\in Z$. Then $a(\tau)\sim
c_+|\tau|^{v_1}+c_-|\tau|^{v_2}.$ Three subcases arise depending
on the value of $G_0$.
\newline
{\bf (RSP1a)} $1/4<G_0$. Then $a(\tau)\sim c|\tau|^{1/2}$,
so that $\lim_{\tau\rightarrow 0^-}a(\tau)=0$..
\newline
{\bf(RSP1b)} $0<G_0<1/4$. Then $v_{1,2}$ are both positive so that
$\lim_{\tau\rightarrow 0^-}a(\tau)=0$.
\newline
{\bf(RSP1c)} $G_0<0$. Then $v_2<0$ so that $\lim_{\tau\rightarrow
0^-}a(\tau)=\infty$.
\newline
{\bf(RSP2)} $v_1=v_2 \Leftrightarrow G_0=1/4$ Then $a(\tau)\sim
c_+|\tau|^{1/2}+c_-|\tau|^{1/2}\ln|\tau|.$ Again,
$\lim_{\tau\rightarrow 0^-}a(\tau)=0$.
\newline
{\bf(RSP3)} $v_1-v_2 \in N-{0} \Leftrightarrow G_0=(1-k^2)/4, k\in
N^+$. Then $a(\tau)\sim
c_+|\tau|^{v_1}+c_-(|\tau|^{v_2}+d|\tau|^{v_1}\ln|\tau|)$ where
$d$ is a {\em fixed} constant. We mention under this last heading
one special case of particular importance, that for which $G_0=0$.
This includes singularities whereat  $F(\tau)$ is finite and,
typically, space-times with weak non-central singularities.
\newline
{\bf (RSP3a)} $G_0=0$. Then $a(\tau)\sim
c_-+c_+|\tau|+c_-d|\tau|\ln|\tau|)$, and so $\lim_{\tau\rightarrow
0^-}a(\tau)$ is finite and non-zero.

The second class of possibilities arises when (\ref{eq15}) has an
irregular singular point at $\tau=0$. If $\lim_{\tau\rightarrow
0^-}|G(\tau)|=\infty$, then the WKB approximation holds. This
gives \[ a(\tau)\sim c(F(\tau))^{-1/4}\exp\{\pm\int_{\tau_1}^\tau
(F(\tau^\prime))^{1/2}\,d\tau^\prime\}.\]  There are two
possibilities here.
\newline
{\bf (ISP1)} $\lim_{\tau\rightarrow 0^-}F(\tau)=+\infty$. Then
$\lim_{\tau\rightarrow 0^-}a(\tau)=+\infty$.
\newline
{\bf (ISP2)} $\lim_{\tau\rightarrow 0^-}F(\tau)=-\infty$. Then
$\lim_{\tau\rightarrow 0^-}a(\tau)=0$.

This does not cover all possibilities since there are irregular
singular points for which the limit $\lim_{\tau\rightarrow
0^-}|G(\tau)|$ does not exist. Typically, this would occur
if $G(\tau)$ is oscillatory in a neighbourhood of $\tau=0$, e.g.
\[G_p(\tau)=k|\tau|^{-1}\sin(|\tau|^{-p}),\] where $p>0$.
Taking this form for $G(\tau)$ and defining $x=|\tau|^{-1}$,
$b=x^{(1+p^{-1})/2}a$ (\ref{eq15}) becomes
\[ b^{\prime\prime}+\left[ \frac{k}{p^2}\frac{\sin
x}{x^{2-p^{-1}}}+\frac{1-p^{-2}}{4x^2}\right] b=0.\] The
dominant coefficient of $b$ is the decaying oscillatory term, and
this determines the asymptotic behaviour of the solutions.
There are three different cases, depending on the value of $p$.
We quote the result for the simplest case, which is $p>1$. The asymptotic behaviour in this case is given by \cite{eastham}
\[ a_1\sim x^{(1-p^{-1})/2},\qquad a_2\sim
x^{-(1+p^{-1})/2}\qquad\qquad (x\rightarrow\infty).\] Thus the
singularity is strong and stretching for $p>1$. It turns out that the singularity is strong and crushing for $1/2<p\leq1$, see \cite{eastham} for details.
On the other hand, if we take $G(\tau)=k\sin(|\tau|^{-1}|)$, the same procedure leads to 
\[ b^{\prime\prime}+\frac{k}{p^2}\frac{\sin x}{x^2}b,\]
the asymptotic solutions of which lead to \cite{eastham}
\[ a_1\sim1,\qquad a_2\sim x^{-1}\qquad\quad (x\rightarrow\infty)\]
and so in this case the singularity is weak. Notice that we have in this case $G(\tau)=O(1)$, but $G$ is not differentiable at $\tau=0$. 

Our main point here is
that both strong and weak singularities may occur in this class
and the analysis to determine which case obtains may be quite
difficult.
\newline
{\bf (ISP3)} $\lim_{\tau\rightarrow 0^-}G(\tau)$ does not exist.
The singularity may be either strong or weak.

Keep in mind that the
behaviour described here is characteristic of a particular radial
timelike geodesic which runs into the singularity, and not of the
singularity itself. We will therefore refer to, for example, a
{\em type (RSP1a) geodesic}, and to a {\em type (RSP1a)
singularity} only if {\em all} the radial timelike geodesics
terminating there are type (RSP1a).

In this language, the central singularity of Schwarzschild
space-time is type (RSP1c), with $a(\tau)\sim
c_+|\tau|^{4/3}+c_-|\tau|^{-1/3}$. Also, $rx(\tau)\sim
x_0|\tau|^{1/3}$, $ry(\tau)\sim y_0|\tau|^{1/3}$, so that overall,
$\|V(\tau)\|\sim d|\tau|^{1/3}$, giving a singularity which is
strong along timelike approaches. Suppose instead the behaviour
was $rx(\tau)\sim x_0|\tau|^{1/6}$, $ry(\tau)\sim
y_0|\tau|^{1/6}$. Then $\|V(\tau)\|\sim d$ (constant), so by the
current definition, the singularity is weak along timelike
approaches. It would be of very little comfort to an observer
jumping into such a singularity to realise, as he watched his legs
elongate and disintegrate, that such volume forms were preserved
on his journey. The possibility of the existence of such a
singularity was noted by Tipler\cite{tipler}. We give an example
of such below. This motivates the following addendum to the
definition of a strong singularity.

We will say that a causal geodesic $\gamma:[\tau_0,0)\rightarrow
M$ approaching a singularity as $\tau\rightarrow0^-$ terminates in
a strong singularity if for all $\tau_1\in[\tau_0,0)$, except some
suitably small set (finite, countable, zero-measure), the general
element of $J_{\tau_1}(\gamma)$ is degenerate or infinite in the
limit $\tau\rightarrow 0^-$. We will say that $\gamma$ terminates
in a weak singularity if the general element of
$J_{\tau_1}(\gamma)$ is finite and non-degenerate in the limit.
The terms will be applied to the singularity itself if {\em all}
causal geodesics approaching the singularity behave in one of the
two ways.

By degenerate, we mean that both of the independent Jacobi fields
in some particular direction (or mutually orthogonal directions)
orthogonal to $k^a$ shrink to zero
magnitude. A non-central type (RSP1b) singularity would be an
example of such.

We now gather the results above into some general statements.

{\bf Proposition Two} {\em For a non-central singularity and for a
central singularity for which ${\dot r}$ has a finite and non-zero
limit along every causal geodesic approaching the singularity, the
strength of the singularity is determined by (\ref{eq12c}). If the
singularity is of type (RSP3a), then it is weak. The only other
type which may be weak is (ISP3); singularities of the other types
are strong. A central singularity for which ${\dot r}$ has zero or
infinite limit  along every causal geodesic approaching the
singularity is strong.}

The proof follows from Proposition One and from the definitions
above; essentially it amounts to some useful book-keeping. A great
many singularities will have different behaviours along different
geodesics approaching the singularity, and so will not be covered
by this result. There remains the problem of determining the
behaviour of $a(\tau)$ and ${\dot r}(\tau)$ in the limit as the
singularity is approached. However we have identified which
elements of the geometry determine the strength of a singularity
and listed the various possibilities.

We now give some applications of the results laid out above.

\section{Examples}

In this section, we study the strengths of some singularities in
four different (classes) of space-times. The first three, two toy
models and Roberts' space-time \cite{Roberts}, are used to
illustrate the types of singularities which may arise and some of
the points made above. The fourth is the marginally bound case of a
Lema\^{\i}tre-Tolman-Bondi (LTB) collapsing dust sphere
\cite{kras}. We use the theory above to demonstrate conclusively
the weakness of shell-crossing singularities in this space-time.

\subsection{A toy model}
We consider the space-time with line-element
\[ ds^2=-dudv+\left(\frac{v-u}{2}\right)^{2\alpha} d\omega^2,\]
i.e. $2e^{-2f}=1$ and $r=((v-u)/2)^\alpha$. We take $\alpha\geq
1$. The case $\alpha=1$ is flat space-time. The Ricci scalar is
\[ R=\frac{2}{r^2}(1-(3\alpha^2-2\alpha)r^{2-2/\alpha}),\]
and so there is a scalar curvature singularity at $r=0$. Since $f$
is constant, the radial Jacobi field orthogonal to an arbitrary
timelike geodesic will satisfy, according to (\ref{eq12c}),
${\ddot a}=0$, with general solution $a=c_++c_-\tau$. Thus the
strength of this central singularity will be determined by the
behaviour of the tangential Jacobi fields.

Along a radial null geodesic, we have (without loss of generality)
$v=$constant and $u=c\tau+d$, where $c,d$ are constants and $\tau$
is an affine parameter. Thus after a reparametrisation of $\tau$,
we have
\[ r=c|\tau|^\alpha.\]
The same result holds for all radial timelike geodesics. From
(\ref{eq12}) and (\ref{eq14}), we find
\[ \|V(\tau)\| \propto (c_+ + c_-|\tau|)|\tau|^{2-2\alpha}.\]
Thus the singularity is strong. Notice that
$\|V(\tau)\|\rightarrow\infty$ as the singularity is approached
along any radial causal geodesic. The deformation results from
infinite tangential stretching.

The purpose of examining this model is to gave an explicit example
where the behaviour at the singularity is clearly pathological and
destructive, but which would not previously have been described as
a strong singularity. We note that
\[ T_{ab}\left(\frac{\partial}{\partial u}\right)^a
\left(\frac{\partial}{\partial u}\right)^b=
-\frac{\alpha}{8\pi}(\alpha-1)r^{-2/\alpha},\]
so that the weak energy condition is violated for the values of $\alpha$
of interest here.

\subsection{Roberts' solution}

Roberts' solution has been used in studies of cosmic censorship
\cite{Roberts} and critical collapse \cite{Brady}. The
line-element is
\[ ds^2=-dudv+\frac{1}{4}(u^2-2uv+(1-p^2)v^2)d\omega^2,\]
where $p$ is constant. $p=0$ gives flat space-time. The Ricci
scalar is
\[ R=\frac{1}{2r^4}p^2uv,\]
and so there is a central scalar curvature singularity. As above,
the strength of the singularity is determined by the tangential
Jacobi fields. In this case we find that $r=c|\tau|$ along any
radial causal geodesic terminating at $r=0$ at parameter value
$\tau=0$. We use (\ref{eq12}) and (\ref{eq14}) to obtain
\[ \|V(\tau)\| \sim k(c_++c_-|\tau|)\]
with $rx,ry\sim$constant as $\tau\rightarrow 0^-$, and so this
central singularity is weak. Thus the examples where this
singularity is naked may not be genuine counterexamples to cosmic
censorship. See also \cite{waldrev} for related comments.

\subsection{Another toy model}
A model with slightly more complicated dynamics and which
illustrates well some of the points made above is that with the
line element
\[
ds^2=-\left(\frac{v-u}{2}\right)^{2\alpha}dudv+
\left(\frac{v-u}{2}\right)^2d\omega^2.\] We take $\alpha\geq 0$;
$\alpha=0$ is flat space-time. We find that \begin{eqnarray}
2e^{2f}f_{uv}=-4\alpha(2\alpha+1)r^{-4\alpha-2},\label{eq16}\end{eqnarray}
and so there is a scalar curvature singularity at $r=0$ (recall
that this term is an invariant).

For an arbitrary radial null geodesic, we make take
$u=u_0$=constant. Then we find
\[ v-u=v-u_0=(c\tau+d)^{1/(2\alpha+1)},\]
so that
\[ r=k|\tau|^{1/(2\alpha+1)}\]
after an appropriate shift in the affine parameter $\tau$.
Applying the second part of Proposition One, we see that all
radial null geodesics approaching $r=0$ terminate in a strong
curvature singularity with the area element obeying
$\|V(\tau)\|\rightarrow 0$ in every case.

To solve for the radial timelike geodesics, we make the change of
variables $r=(v-u)/2$, $t=(v+u)/2$. Then the line element takes
the form
\[ ds^2=r^{2\alpha}(-dt^2+dr^2)+r^2d\omega^2.\]
The geodesic equations for radial infall yield
\[ {\dot r}=-r^{-2\alpha}(c^2-r^{2\alpha})^{1/2}.\]
According to (\ref{eq12c}) and (\ref{eq15}), we need to determine
the behaviour of $r$ as proper time $\tau\rightarrow 0$. (As
usual, we fix the origin of proper time so that $r(0)=0$.) The
previous equation may be solved asymptotically by expanding the
right hand side and then inverting the resulting integral with the
result \begin{eqnarray} r&=&c|\tau|^{1/(2\alpha+1)}+O(|\tau|^\beta)
\label{eq17}\end{eqnarray}where $\beta>1/(2\alpha+1)$.

Then the tangential Jacobi fields have the asymptotic behaviour
\begin{eqnarray*} rx&\sim&c|\tau|^{2\alpha/(2\alpha+1)}\qquad (\alpha\neq
1/2),\\ rx&\sim&c|\tau|^{1/2}\ln|\tau|\qquad (\alpha=1/2).
\end{eqnarray*}

The behaviour of the radial Jacobi fields is dictated by
(\ref{eq16}) which from the above has the behaviour
\[ F(\tau)\sim c_1|\tau|^{-2}\]
where $c_1$ is a {\em negative} constant. Then in the notation
used above, $G_0$ is a negative constant, and so this is a type
(RSP1c) singularity. The asymptotic behaviour of $a$ is
\[ a(\tau)\sim c_+|\tau|^{v_1}+c_-|\tau|^{v_2} \]
where $v_{1,2}=(1\pm(1-4G_0)^{1/2})/2$, and so
\[ \|V(\tau)\|\sim V_0|\tau|^{v_2+4\alpha/(2\alpha+1)}.\]
Therefore, for any value of $\alpha$, there will exist radial
timelike geodesics along which $\|V(\tau)\|$ diverges, has zero
limit and has finite limit as the singularity is approached. These
different possibilities arise from the different choices available
for $c$ in (\ref{eq17}) which give the value
$G_0=-4\alpha(2\alpha+1)c^{-4\alpha-2}$. Starting from some fixed
value $r=r_*$ at $\tau=\tau_*<0$, we see that $c$ is essentially a
measure of the initial velocity of an observer falling radially
inwards from $r_*$. By tuning this velocity, an observer could in
principle ensure that his $\|V(\tau)\|$ is finite in the approach
to the singularity. However, in practice this would be of little
help to him since as pointed out above, the observer experiences
infinite tangential crushing and radial stretching in the infall.
Furthermore, his initial velocity would have to be tuned with
infinite precision to obtain $0\neq\lim_{\tau\rightarrow
0^-}\|V(\tau)\|<\infty$. According to the definition above, this
is a strong singularity.

\subsection{Marginally bound spherical dust}

The marginally bound LTB space-time (spherically symmetric
inhomogeneous dust) has line element
\[ ds^2=-dt^2+(r^\prime)^2d\eta^2+r^2d\omega^2,\]
where the prime indicates differentiation with respect to the
coordinate $\eta$. For the collapsing case,
\[ r^3(\eta,t)=\frac{9}{2}m(\eta)(t_0(\eta)-t)^2,\]
where $m,t_0$ are arbitrary functions of $\eta$. See \cite{kras}
for details. The energy density $\rho$ of the dust, which is
proportional to the Ricci scalar, is given by
\[ 4\pi\rho =\frac{m^\prime}{r^2r^\prime}.\]
Thus as well as the central singularity at $r=0$ (occurring when
$t=t_0(\eta)$, there are so-called shell-crossing singularities
occurring when $r^\prime=0$ \cite{yod}. These generally occur
before the central singularity, at non-zero radius and so are
non-central. It has long been believed that these scalar curvature
singularities are weak. However, it seems that this has only been
properly established for null geodesics approaching the
singularity \cite{joshi,newman}. As we have seen above, this
weakness is completely independent of the structure and nature of
the singularity apart from the fact that it is non-central. We
fill this gap by proving that all radial timelike geodesics
terminate in a weak singularity.

According to Proposition Two, the strength of a shell-crossing
singularity is governed by (\ref{eq12c}). Using (\ref{eq5}), we
find that
\[ F=2e^{2f}f_{uv}=\frac{m^\prime}{r^2r^\prime}-2\frac{m}{r^3}.\]
The terms $m^\prime/r^2$ and $m/r^3$ will both be finite in
general in the approach to the singularity, the former being
positive, assuming positive energy density. Thus the behaviour is
governed by $F=1/r^\prime$. We will show that \begin{eqnarray}
\lim_{\tau\rightarrow 0^-} \frac{\tau^2}{r^\prime}&=&0
\label{eq18}\end{eqnarray}along any radial timelike geodesic approaching the
singularity. This shows that the singularity is type (RSP3a), and
is therefore weak by Proposition Two.

The radial timelike geodesic equations are
\begin{mathletters}\label{eq19}
\begin{eqnarray}
-{\dot t}^2+(r^\prime)^2{\dot\eta}^2&=&-1 \label{eq19a}\\
r^\prime{\ddot \eta}+2r^\prime_t{\dot t}{\dot
\eta}+r^{\prime\prime}{\dot \eta}^2&=&0 \label{eq19b}\\ {\ddot
t}+r^\prime r^\prime_t{\dot \eta}^2&=&0\label{eq19c} \end{eqnarray}
\end{mathletters}
where the overdot indicates differentiation with respect to proper
time along the geodesic and the subscript is differentiation with
respect to the global time coordinate $t$. Along each geodesic
approaching the singularity, we choose the origin of proper time
so that the singularity is at $\tau=0$.

We find that
\[ r^\prime
=\frac{r}{3}\left(\frac{m^\prime}{m}+\frac{2t_0^\prime}{t_0-t}\right),\]
so that at a shell-crossing singularity,
\[ \frac{m^\prime}{m}=-\frac{2t_0^\prime}{t_0-t}.\]
The following terms will enter into our analysis;
\begin{eqnarray}
r^\prime_t&=&\frac{2}{3}\frac{rt_0^\prime}{(t_0-t)^2}
\label{eq20}\\ r^{\prime\prime}&=&\frac{(r^\prime)^2}{r}+
\frac{r}{3}\left(\frac{m^{\prime\prime}}{m}-\left(\frac{m^\prime}{m}\right)^2
+\frac{2t_0^{\prime\prime}}{t_0-t}-\frac{2t_0^\prime}{(t_0-t)^2}\right).
\label{eq21}
\end{eqnarray}
Then \begin{eqnarray} r^{\prime\prime}(0)&=&\frac{r}{3m^2t_0^\prime}
\left(mm^{\prime\prime}t_0^\prime-
\frac{3}{2}(m^\prime)^2t_0^\prime-mm^\prime
t_0^{\prime\prime}\right),\label{eq22}\end{eqnarray}where here and
subsequently, evaluation at zero means in the limit
$\tau\rightarrow 0^-$ along a geodesic.

Generically, $r^\prime_t(0)$ and $r^{\prime\prime}(0)$ will be
non-zero. If this were not the case, there would be extra
conditions imposed on $m$ and $t_0$ for all values of $\eta$,
which would result in a loss of generality. For example, if
$r^\prime_t(0)=0$, then $t_0^\prime(\eta)\equiv0$ for all $\eta$.
In this case, the space-time is homogeneous and isotropic. The
condition $r^{\prime\prime}(0)=0$ imposes less severe but
nonetheless significant restrictions. So we assume henceforth that
$r^\prime_t(0)$ and $r^{\prime\prime}(0)$ are non-zero.

We also need to track the evolution of $r$ and $r^\prime$ along
the geodesics. We have \begin{eqnarray} {\dot r}&=& r_t{\dot t}+r^\prime{\dot
\eta}=\sqrt{\frac{2m}{r}}{\dot t}+r^\prime{\dot \eta}
\label{eq23}\\ {\dot
{(r^\prime)}}&=&\frac{2}{3}\frac{rt_0^\prime}{(t_0-t)^2}+r^{\prime\prime}{\dot
\eta}.\label{eq24}\end{eqnarray}

We now prove (\ref{eq18}), which demonstrates the weakness of the
singularity.

{\bf Case One:}$\lim_{\tau\rightarrow 0^-}|{\dot\eta}|<\infty|$.

By (\ref{eq19a}), ${\dot t}(0)=1$. The sign comes from the
assumption that the geodesic is future directed and the fact that
$t$ is a global time coordinate. The past directed case proceeds
in an identical manner. By (\ref{eq24}), ${\dot{(r^\prime)}}$ will
be finite in the limit $\tau\rightarrow 0^-$. If this limit is
non-zero, we can apply l'Hopital's rule to $\tau^2/r^\prime$ to
prove (\ref{eq18}). The other possibility is that
${\dot{(r^\prime)}}(0)=0$ So now assume this to be the case.

Suppose further that $|{\ddot \eta}(0)|<\infty$. Then $(r^\prime
{\ddot \eta})(0)=0$, and so taking the limit of (\ref{eq19b}), we
have
\begin{eqnarray*}
0&=&\lim_{\tau\rightarrow 0^-} (2r^\prime_t {\dot
t}+r^{\prime\prime}{\dot \eta})\\ &=& \lim_{\tau\rightarrow
0^-}(r^\prime_t{\dot t}+{\dot{(r^\prime)}}),
\end{eqnarray*}
which gives $r^\prime_t(0)=0$, in contradiction of one of our
assumptions. So if ${\dot{(r^\prime)}}(0)=0$, then we must have
$|{\ddot \eta}(0)|=\infty$.

Using l'Hopital's rule twice, we have in this case
\[ \lim_{\tau\rightarrow 0^-}\frac{\tau^2}{r^\prime}=
\lim_{\tau\rightarrow 0^-}\frac{2}{{\ddot{(r^\prime)}}}.\] We
calculate
\[ {\ddot{(r^\prime)}}= r^\prime_{tt}{\dot t}^2 +r^\prime_t{\ddot
t}+2r^{\prime\prime}_t{\dot t}{\dot \eta}+r^{\prime\prime}{\ddot
\eta}+r^{\prime\prime\prime}{\dot \eta}^2.\] From (\ref{eq19c}),
${\ddot t}(0)=0$ and the terms $r^\prime_{tt}$,
$r^{\prime\prime}_t$ and $r^{\prime\prime\prime}$ will be finite in
the appropriate limit. Thus the dominant term is
$r^{\prime\prime}{\ddot \eta}$, giving $\lim_{\tau\rightarrow
0^-}|{\ddot{r^\prime}}|=\infty$, proving (\ref{eq18}).

{\bf Case Two:} $\lim_{\tau\rightarrow 0^-}|{\dot{\eta}}|=\infty$.

Suppose that $|r^\prime{\dot \eta}|(0)<\infty$. Then by
(\ref{eq19a}), ${\dot t}(0)$ is finite and so (\ref{eq24}) gives
$|{\dot{(r^\prime)}}|(0)=\infty$. We then use l'Hopital's rule to
prove (\ref{eq18}.

Finally, suppose that $|r^\prime{\dot \eta}|(0)=\infty$. Then by
(\ref{eq19a}),
\begin{eqnarray*}
\lim_{\tau\rightarrow 0^-} {\dot t}&=&\lim_{\tau\rightarrow
0^-}\left(|r^\prime{\dot \eta}|\left(1+\frac{1}{|r^\prime{\dot
\eta}|}\right)^{1/2}\right)\\ &=&\lim_{\tau\rightarrow
0^-}|r^\prime{\dot \eta}|.
\end{eqnarray*}
Then by (\ref{eq24}), \[ \lim_{\tau\rightarrow
0^-}|{\dot{(r^\prime)}}|=|r^{\prime\prime}\pm r^\prime
r^\prime_t|(0) \lim_{\tau\rightarrow 0^-}|{\dot\eta}|.\] This can
be finite only if $(r^{\prime\prime}\pm r^\prime
r^\prime_t)(0)=0$. But this limit is generically equal to
$r^{\prime\prime}(0)$, which is non-zero, and so we have
$|{\dot{(r^\prime)}}|(0)=\infty$. Again, l'Hopital's rule is used
to prove (\ref{eq18}).

This completes the proof of (\ref{eq18}) for all radial timelike
geodesics and thus demonstrates the weakness of the singularity.

\section{Conclusions}
The central results here are contained in equations (\ref{eq12}),
(\ref{eq12c}) and (\ref{eq14}). These provide a set of covariant
equations, the asymptotic solutions to which (which require
information about causal geodesics) determine the strengths of
singularities in spherically symmetric space-times. The notion of
`strength' is in a slightly modified form to Tipler's original
definition \cite{tipler}; the modification is clearly motivated
and is illustrated by the examples in Section 3.

Proposition One demonstrates the important point that the
behaviour of null geodesics tells us nothing about the strength of
a non-central singularity. Also, a null geodesic approaching a
central singularity terminates in a strong singularity unless
${\dot r}$ has a finite, non-zero limit at the singularity.
Proposition Two lists the possible ways in which strong or weak
singularities may occur.

In addition to studying the toy models, we were able to
demonstrate conclusively the weakness of the naked singularity in
Roberts' space-time and the shell-crossing singularities in
collapsing spherical dust. This latter proof shows that while
detailed qualitative information about causal geodesics is
required, we do not need the full solution of the geodesic
equations. Therefore there is good hope that the results above may
be successfully applied to other situations.

One of these is the case where the singularity occurs at a point
where the metric is continuous and non-degenerate
($\det(g_{ab})\neq 0$) (we will refer to such as a continuous non-degenerate singularity). It seems plausible that in such a
situation, the singularity must necessarily be weak. The argument
goes roughly as follows. Solutions of the time-like geodesics of
section 2 typically behave as ${\dot{u}}=O(1)$, ${\dot {v}}=O(1)$
as $\tau\rightarrow 0$. Then to obtain a strong curvature
singularity, the Riemann tensor components must diverge faster
than $O(\tau^{-2})$; integrating twice cannot yield a finite
metric. However this argument might not hold for an (ISP3)
singularity, and perhaps not for other cases. A careful analysis of (\ref{eq6}) and (\ref{eq12c})
should be able to yield either a theorem stating that a
continuous non-degenerate singularity is indeed weak, or produce
examples to the contrary. The statement that a continuous non-degenerate singularity is necessarily weak has been made, or the conclusion been used, on several occasions in the literature in connection with studies of the Cauchy horizon singularity in black holes and singularities in plane wave space-times. This has usually been accompanied by separate calculations verifying that the singularity is indeed weak \cite{ori92,oriburkop19}, but this has not always been the case \cite{hod-piran,burko,oripw}. Thus it appears to be of importance to determine exactly when one can conclude weakness for a continuous non-degenerate singularity.  

Clarke and Krolak \cite{CK} have given necessary and sufficient
conditions, in arbitrary space-times, for a singularity to be
strong, the conditions involving integrals of certain curvature
terms along geodesics. An advantage of our work is that it deals
with the full set of Jacobi fields $J_\tau$ rather than the volume
element $V(\tau)$. As the toy model of Section 3.3 shows, this can
be important. Also, the decisive term here $2e^{2f}f_{uv}$ is
slightly simpler than the decisive terms in \cite{CK}. It may be
possible to use the results here to investigate the connection
between Tipler's definition of strengths of singularities and
Krolak's limiting focusing conditions \cite{krolak}.

\section*{Acknowledgements}
I am grateful to Shahar Hod for bringing the first of 
ref. \cite{hod-piran} to my attention and to Alastair Wood for suggesting ref. \cite{eastham}.

\end{document}